\documentclass[a4paper,onecolumn,preprint,review,12pt,amsmath,amssymb,sort&compress,preprintnumbers]{elsarticle}
\usepackage{graphicx}
\usepackage{epsfig}
\usepackage{multirow}
\usepackage{amssymb}
\usepackage{amsmath}
\usepackage{color}
 \usepackage{lineno}
 \usepackage{geometry}
 \usepackage{lipsum}
\journal{arXiv}
\begin{document}
\begin{frontmatter}
\newcommand\blfootnote[1]{%
\begingroup
\renewcommand\thefootnote{}\footnote{#1}%
\addtocounter{footnote}{-1}%
\endgroup
}
\title{Hydrogen bubble nucleation by self-clustering: Density Functional Theory and statistical models studies using tungsten as a model system}
\author{Jie Hou$^{a, b, c,1}\blfootnote{1 These authors contributed equally to this work}$, Xiang-Shan Kong$^{a,1,*} \blfootnote{*Author to whom correspondence should be addressed. Email address: xskong@issp.ac.cn}$, Jingjing Sun$^{a, b}$, Yu-Wei You$^{a}$, Xuebang Wu$^{a}$, C. S. Liu$^{a,**}\blfootnote{**Author to whom correspondence should be addressed. Email address: csliu@issp.ac.cn Tel: 0086-551-65591062}$, Jun Song$^{c,***}\blfootnote{***Author to whom correspondence should be addressed. Email address: jun.song2@mcgill.ca}$}
\address{$^{a}$Key Laboratory of Materials Physics, Institute of Solid State Physics, Chinese Academy of Sciences, P. O. Box 1129, Hefei 230031, P. R. China

$^b$ University of Science and Technology of China, Hefei 230026, P. R. China

$^c$ Department of Mining and Materials Engineering, McGill University, Montr$\acute{e}$al, Qu$\acute{e}$bec H3A 0C5, Canada}

\begin{abstract}
%% Text of abstract
{\color{black}Low-energy hydrogen irradiation is known to induce bubble formation in tungsten, while its atomistic mechanisms remain little understood. Using first-principles calculations and statistical models, we studied the self-clustering behavior of hydrogen in tungsten. Unlike previous speculations that hydrogen self-clusters are energetically unstable owing to the general repulsion between two hydrogens, we demonstrated that  hydrogen self-cluster becomes more favorable as the cluster size increases. We found that hydrogen atoms would form two-dimensional platelet-like structures along $\{100\}$ planes. These hydrogen self-clustering behaviors can be quantitative understood by the competition between long-ranged elastic attraction and local electronic repulsion. Further statistical analysis showed that there exists a critical hydrogen concentration above which hydrogen self-clusters are thermodynamically stable and kinetically feasible. Based on this critical hydrogen concentration, the plasma loading conditions under which hydrogen self-clusters form were predicted. Our predictions showed excellent agreement with experimental results of hydrogen bubble formation in tungsten exposed to low-energy hydrogen irradiation. Finally, we proposed a possible mechanism for the hydrogen bubble nucleation via hydrogen self-clustering. This work provides mechanistic insights and quantitative models towards understanding of plasma-induced hydrogen bubble formation in plasma-facing tungsten.}

\end{abstract}

\end{frontmatter}
\section{Introduction}
 Hydrogen (H) bubble formation, and the consequent hydrogen embrittlement it often results, posts a great threat to the structural integrity and mechanical properties of metals, promoting extensive studies of it over the years.\cite{Xie2015,Tiegel2016,Rozenak2007,Condon1993} Several mechanisms, including "loop punching" and "vacancy clustering", have been proposed to explain the hydrogen bubble growth.\cite{Condon1993} While these mechanisms are quite distinct in nature, they all require a nucleation process for the respective hydrogen bubble growth to occur. To date, it is generally believed that hydrogen bubble nucleation by self-clustering like the case of helium in metals \cite{Becquart2006,Kong2016-2} would be impossible given the H-H strong repulsion or very weak attraction in metals \cite{Kong2016-1,Henriksson2005,Becquart2009,Liu2009,You2014,Monasterio2009,Ouyang2011,Pezold2011}, and that, as suggested by many previous studies \cite{Condon1993,Henriksson2005,Kong2016-1,Liu2009,You2014}, hydrogen bubble nucleation would require the presence of lattice defects. The hydrogen bubble nucleation can be either heterogeneous, relating to grain boundaries,\cite{Zhou2016,Valles2017,Zhou2010,Xiao2012} dislocations,\cite{Xiao2012} and impurities,\cite{Kong2016-1,Kong2013} or homogeneous, arising from the aggregation of vacancies or vacancy-hydrogen complexes.\cite{Geng2017,Sun2016} However, hydrogen bubble formation in metals with extremely low concentration of the lattice defects has been clearly observed in many experiments. \cite{Tetelman1963,Ren2008,Escobar2011,Poon2005,Alimov2005,Alimov2008,Poon2002,wang2001,Shu2007-2,Shu2007-1,Shu2008,Alimov2009,Alimov2012,Luo2005,Zibrov2017,Tokunaga2005,Zayachuk2015,Ni2015,Buzi2014,Buzi2015,Jia2017} In particular, one most typical phenomenon over the past decade is hydrogen bubble formation in annealed tungsten exposed to high-flux low-energy hydrogen irradiation.\cite{Poon2005,Alimov2005,Alimov2008,Poon2002,wang2001,Shu2007-2,Shu2007-1,Shu2008,Alimov2009,Alimov2012,Luo2005,Zibrov2017,Tokunaga2005,Zayachuk2015,Ni2015,Buzi2014,Buzi2015,Jia2017}

 Since tungsten has been considered as the primary candidate for the plasma-facing material (PFM) of fusion reactors, hydrogen bubble formation in tungsten received great attention. Hydrogen bubble formation in tungsten not only leads to instability of the plasma but also increases tritium inventory in the PFM.\cite{Shu2007-2} Despite extensive research studies, it is still an open question how hydrogen bubbles nucleate during low-energy and high-flux hydrogen plasma irradiations. On one hand, the low energy of hydrogen plasma used in the experiments is far lower than the threshold energy needed to induce considerable atom displacement in tungsten for the formation of new self-interstitial atoms or vacancies. On the other hand, in most of those previous experiments \cite{Alimov2005,Poon2002,wang2001,Shu2007-2,Shu2007-1,Shu2008,Alimov2009,Alimov2012,Luo2005,Zibrov2017,Tokunaga2005,Zayachuk2015,Ni2015} annealed pure tungsten samples with coarse-grained or single crystal structure were used, which minimizes the influence of the intrinsic defects, such as dislocations, vacancies, and grain boundaries, produced during the sample preparation. Furthermore, the formation energies of lattice defects in tungsten are very high. For instance, even for the simple point defect, i.e., a single vacancy, the formation energy is about 3.2 eV \cite{Kong2013}, which corresponds to a very low vacancy concentration (about $10^{-54}$, in atomic ratio, similarly hereinafter) at room temperature and even at the annealing temperature (e.g., about $10^{-13}$ at 1270 K). Therefore, population of lattice defects in well-annealed undamaged tungsten is expected to be rather limited, being insufficient to provide enough nucleation sites for hydrogen bubble formation.

 The afore-mentioned puzzle regarding hydrogen bubble nucleation stimulates many dedicated studies. Poon et al. argued that the plasma impurities, such as carbon and oxygen, have energy transfer efficiencies higher than that of hydrogen isotopes, therefore can create vacancies more easily to supply nucleation sites when tungsten was exposed to a 500 eV deuterium plasma.\cite{Poon2005} However, this mechanism cannot be applied to other experiments where hydrogen bubbles were also observed after exposed to deuterium plasma with energy of tens of eV.
 \cite{Shu2007-2,Shu2007-1,Shu2008,Alimov2009,Alimov2012,Luo2005,Zibrov2017,Tokunaga2005,Zayachuk2015,Ni2015,Buzi2014,Buzi2015,Jia2017} To explain these experiments \cite{Shu2007-2,Shu2007-1,Shu2008,Alimov2009,Alimov2012,Luo2005,Zibrov2017,Tokunaga2005,Zayachuk2015,Ni2015,Buzi2014,Buzi2015,Jia2017}, Shu et al. employed Fukai's theory of superabundant vacancies (SAV)\cite{Shu2007-1,Shu2007-2}, attributing the high vacancy concentration to the significant reduction of the vacancy formation energy by vacancy-H complexes.\cite{Fukai2003,Sugimoto2014,Sugimoto2017} Nonetheless, a recent theoretical study by Kong et al.\cite{Kong2013} showed that vacancy-H complexes in tungsten can only reduce the vacancy formation energy to $\sim2.45$ eV, being insufficient to produce a high vacancy concentration ($\sim 10^{-39}$ at room temperature). Furthermore, the SAV formation often requires high temperatures, which are not provided by many hydrogen bubble formation experiments.\cite{Poon2005,Alimov2005,Alimov2008,Poon2002,wang2001,Shu2007-2,Shu2007-1,Shu2008,Alimov2009,Alimov2012,Luo2005,Zibrov2017,Tokunaga2005,Zayachuk2015,Ni2015} Kong et al. showed that oxygen impurities can significantly decrease the vacancy formation energy in tungsten and suggested that the vacancy concentration may be enhanced by the formation of vacancy-oxygen-hydrogen complexes. \cite{Kong2013} However, hydrogen nano-bubbles were only observed in the subsurface region in recent experiments, indicating no obvious relationship of hydrogen bubble formation with the impurities.\cite{Jia2017}The nucleation mechanisms proposed by the above studies are related to the vacancy formation. Though they have their individual merits and may operate under certain conditions, they are not general and fundamental enough to provide quantitative explanations of hydrogen bubble nucleation in tungsten exposed to low-energy and high-flux hydrogen plasma irradiations. Moreover, hydrogen bubble formation was frequently observed at around room temperature. \cite{Alimov2005,Alimov2008,Poon2002,wang2001,Shu2007-2,Shu2007-1,Shu2008,Alimov2009,Alimov2012,Luo2005,Zibrov2017,Tokunaga2005,Zayachuk2015,Ni2015}, at which the vacancies or vacancy-hydrogen complexes, even if they exist with high concentrations, would be immobile and thus not able to gather together to form bubbles.

 Recently, by analyzing the spatial distribution of low energy hydrogen induced defects, Ni et al. \cite{Ni2015} and Jia et al. \cite{Jia2017} find that the hydrogen bubble distribution is homogeneous and related to neither original vacancies nor impurities. These results suggest the existence of a possible hydrogen bubble nucleation mechanism independent of pre-exist lattice defects. Alimov et al. have proposed a void formation mechanism to explain the sudden rise in deuterium trapping sites and the concurrent deuterium accumulation in tungsten exposed to 200 eV deuterium ions.\cite{Alimov2005} They suggested that excessive interstitial deuterium in the implantation zone severely stresses the tungsten lattice, and subsequently causes formation of voids or vacancy clusters to alleviate these tensions. Unfortunately, no atomic detail has been given to support this suggestion. In addition, it is worth noting that some early experimental and theoretical researches seem to suggest the aggregation of hydrogen in a platelet shape in metals, which can alleviate the strain induced by the excessive interstitial hydrogen atoms.\cite{Fujita1976,Kamachi1972} {\color{black}In this work, we focus on the self-clustering behavior of hydrogen, i.e., hydrogen clustering in a defect-free tungsten lattice. First-principles calculations were carried out to explore possible hydrogen self-cluster structures and to identify the most favorable ones. We found that hydrogen self-clusters prefer to form planar structures within $\{100\}$ planes. We show that self-clustering behavior can be well described by the competition between elastic attraction and electronic repulsion. Based on the planar hydrogen self-clustering, we propose a bubble formation mechanism in defect-free tungsten. Further statistic evaluations show that the self-clustering is thermodynamically favorable and kinetically feasible at high hydrogen concentrations. Our results provide quantitative explanation for previous experiments where bubbles were induced by low energy hydrogen plasma.}
\section{Computation method}
First-principles calculations on the basis of density functional theory (DFT) as implemented in the Vienna ab-initio simulation package (VASP) \cite{Kresse1993,Kresse1996} with Blochl's projector augmented wave (PAW) potential method \cite{Blochl1994} were performed. All 5d and 6s electrons of tungsten and the 1s electron of hydrogen were treated as valence electrons in the PAW potential. The generalized gradient approximation and the Perdew-Wang function were used to describe the electronic exchange and correlation effect\cite{Perdew1992,Perdew1993}. A supercell consisting of 288 lattice points (a $6\times6\times4$ duplicate of a conventional bcc unit cell) was used to mimic bulk conditions. Relaxations of atomic configuration and optimizations of the shape and size of the super-cell were performed for all calculations unless stated otherwise. A plane wave cutoff of 500 eV and \emph{k}-point density of $2\times2\times3$, obtained using the Monkhorst-Pack method, were used in these calculations. Benchmark calculations with increased super-cell size, cutoff energy and k-point density, have been carried out, and negligible influence on our results was found. The convergence criteria of system energy and atomic force were set as 1 $\mu$eV and 0.01 eV/{\AA} respectively in our calculations.

 The binding energy of an isolated interstitial hydrogen atom with a stable $H_{n-1}$ self-cluster is calculated as:
 \begin{equation}
 E_{b}^{H_n}=E^{H_{n-1}}_{tot}+E^{H_1}_{tot}-(E^{H_n}_{tot}+E_{tot}^{bulk}),
 \end{equation}
 where $E^{H_n}_{tot}$ and $E_{tot}^{bulk}$ are energies of super-cells with and without a $H_n$ self-cluster. A negative value of the binding energy indicates repulsion between two defects, while the positive value means attraction. {\color{black}The solution energy of an isolated interstitial hydrogen atom in tungsten is calculated as:
 \begin{equation}
 E_{s}= E^{H_1}_{tot}-E_{tot}^{bulk}-\frac{1}{2}E^{H_{2}},
 \end{equation}
 where $E^{H_{2}}$ is the energy of a $H_2$ molecule in vacuum.

 In addition, the zero-point energy (ZPE) correction, which can play an important role for light elements, was also included in this work to ensure accurate description of energies of hydrogen and hydrogen self-clusters. The binding energy and solution energy with ZPE corrections are given by:
 \begin{equation}
 E_{b}^{H_n}(ZPE)= E_{b}^{H_n}+E^{H_{n-1}}_{ZPE}+E^{H_1}_{ZPE}-E^{H_n}_{ZPE},
 \end{equation}
 \begin{equation}
 E_{s}(ZPE)=E_{s}+E^{H_1}_{ZPE}-\frac{1}{2}E^{H_{2}}_{ZPE},
 \end{equation}
 where $E^{H_n}_{ZPE}$ is total zero-point energy of the $H_n$ self-cluster, and $E^{H_{2}}_{ZPE}$ is the total zero-point energy of a $H_2$ molecule in vacuum. The zero-point energy for the tungsten atom is usually negligible, and thus not considered here for computational efficiency.}
\section{Results and discussion}
\subsection{Configurations and binding energetics of stable $H_n$ self-clusters}
\begin{figure}[h]
 \begin{center}
 \includegraphics[width=14cm]{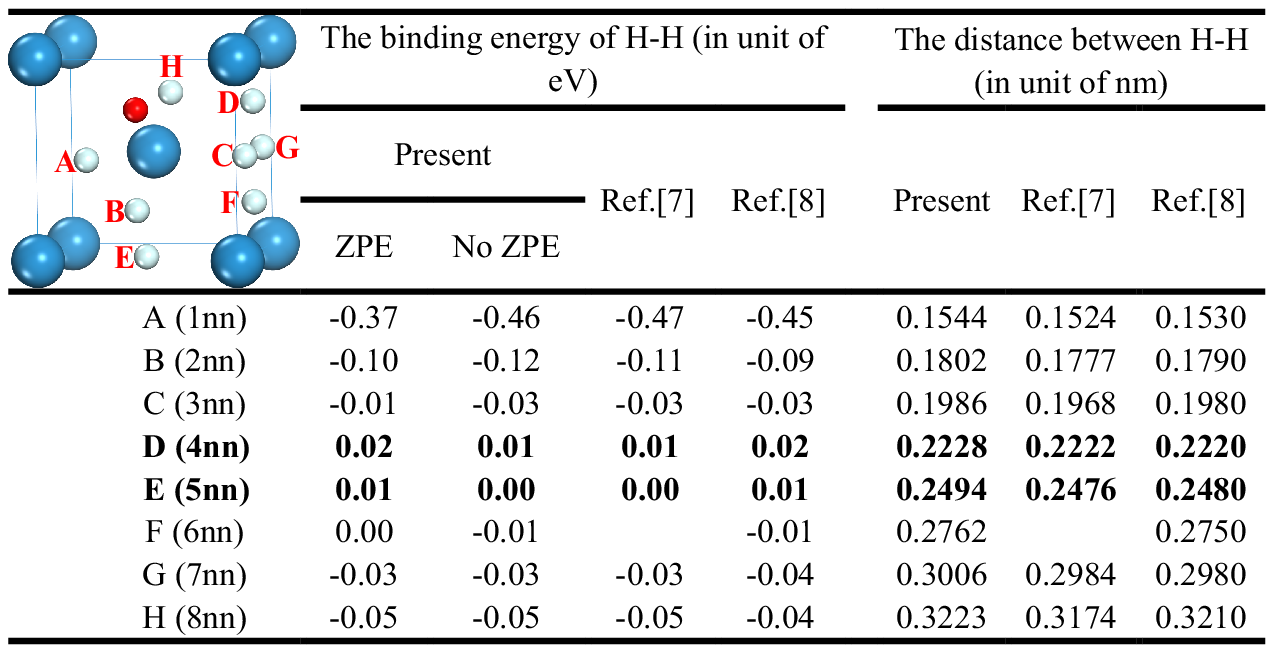}
 \end{center}
 \caption{(color online) The binding energies between two hydrogen atoms located at 1nn to 8nn neighboring TIS sites, calculated using Eqs. (1) and (3). The insert figure shows spatial relationship between two hydrogen atoms. The large mineral blue balls are tungsten atoms. The small red ball marks out position of the reference hydrogen atom while the small white balls, denoted by letters A$\sim$H, indicate positions of the second hydrogen atom at 1nn$\sim$8nn neighboring TIS sites. }
 \end{figure}

 It is well known that in tungsten an interstitial hydrogen atom prefers to occupy the tetrahedral interstitial site (TIS) rather than the octahedral interstitial site (OIS), and the interstitial H-H interaction is generally repulsive.\cite{Kong2016-1,Henriksson2005,Becquart2009,Liu2009,You2014} These have also been confirmed by our current calculations where the TIS is found to be 0.38 eV more stable than the OIS for a hydrogen atom. {\color{black}Our preliminary calculations also show that all TIS-OIS hydrogen pairs, OIS-OIS hydrogen pairs, and OIS hydrogen self-clusters are energetically unfavorable or unstable compared with the pure TIS ones. Therefore, we only consider TIS hydrogens in this study.} Figure 1 summarizes the binding energies between two hydrogen atoms located at $1nn$ to $8nn$ neighboring TIS sites, showing that the binding energy of H-H pairs is strongly negative ($\sim$-0.46 eV) at the distance of $\sim$1.54 {\AA}, then increases rapidly with the increase of the H-H separation distance, and becomes a small positive value ($\sim$0.01 eV) at a distance of $\sim$2.22 {\AA}. With further increasing distance, the binding energy decreases into a small negative value again before it eventually diminishes to zero. These results suggest that two interstitial hydrogen atoms in tungsten be overall repulsive. Our results are also in good agreement with results previously reported \cite{Kong2016-1,Henriksson2005,Becquart2009,Liu2009,You2014}. However it is worth to note two particular cases of H-H pairing, one being two hydrogen atoms located at the 4nn neighboring TIS sites to form a H-H pair along the $<$110$>$ direction, and the other being two located at the 5nn neighboring TIS sites to form a H-H pair along the $<$310$>$ direction, both yielding positive binding energies. Despite the attractive interaction being weak, it hints a possibility of forming hydrogen self-clusters along particular directions or within particular planes. Moreover, some previous studies on inert-gas self-clusters in tungsten suggested that attraction between interstitial atoms may become stronger as the local aggregation grows \cite{Becquart2006,Kong2016-2}, eventually leading to large self-clusters. In analogy to those studies, it is important to investigate if a similar scenario can occurs for H-H interaction to finally promote the formation of sizable hydrogen self-clusters, as elaborated in the follows.

 \begin{figure}[htp]
 \begin{center}
 \includegraphics[width=15cm]{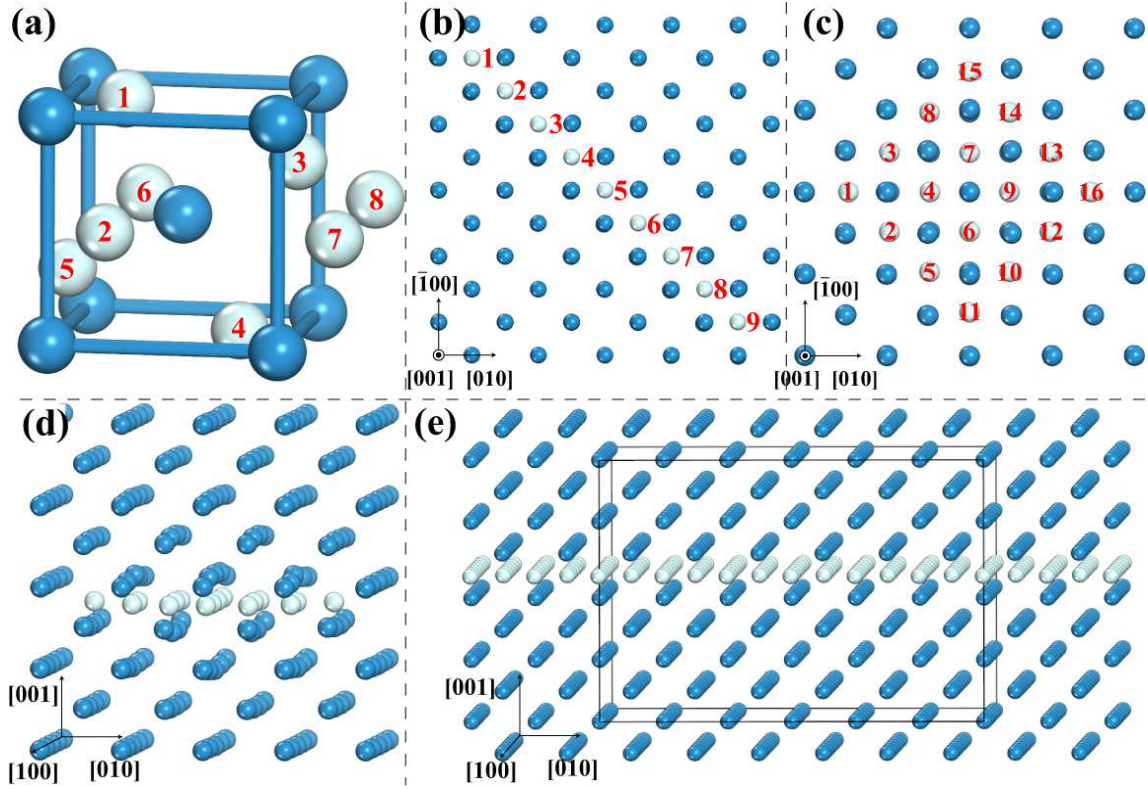}
 \end{center}
 \caption{(color online) Configurations of hydrogen self-clusters. Tungsten and hydrogen atoms are in blue (balls/lines) and white respectively. The red numbers represent hydrogen occupation sequence. (a) and (b) illustrates the most stable spherical hydrogen self-cluster and linear hydrogen self-cluster along [110] direction respectively. (c) and (d) respectively show the top and side views of the most stable planar $H_n$ self-cluster in a (001) plane. (e) shows the planar $H_{n\rightarrow\infty}$ self-cluster  in a periodic supercell.}
 \end{figure}

 \begin{figure}[htp]
 \begin{center}
 \includegraphics[width=7cm]{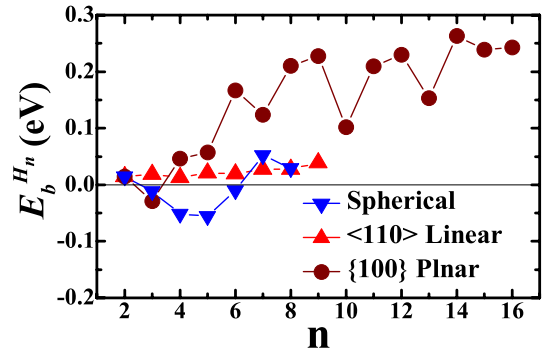}
 \end{center}
 \caption{(color online) Binding energies of hydrogen self-clusters (i.e., those previously shown in Figure 2) as the cluster size $n$ varies.}
 \end{figure}

 In order to check the possibility of interstitial hydrogen self-cluster ($H_n$) formation, we first constructed a series of plausible geometries. A hydrogen atom is introduced one by one to the TISs neighboring the stable and metastable $H_{n-1}$ ($n>2$) complexes and then the total energy is minimized to identify the most stable configuration of the resultant $H_n$ self-clusters. However, because of the overwhelming number of possible combinations of the $H_{n-1}$ self-cluster and additional hydrogen, it is impossible to screen all possible configurations. To overcome this challenge, we utilize two criteria to select possible positions for an additional hydrogen atom introduced to the vicinity of the $H_{n-1}$ cluster, accounting for the energetics of H-H pair interaction: 1) the additional hydrogen atom, when introduced, forms the the $<$110$>$ or $<$310$>$ H-H pairs with those original hydrogen atoms within $H_{n-1}$ as many as possible; 2) the distance between the newly introduced hydrogen and an original hydrogen within $H_{n-1}$ is larger than 0.2 nm ($\ge$3nn). \cite{Switendick1979}

 {\color{black} Using the above strategy, we systematically investigated hydrogen self-clusters in three different shapes: spherical, linear, and planar. Figure 2 illustrates the energetically preferable configurations for each of those self-cluster shapes. The corresponding binding energies as functions of the cluster size $n$ are shown in Figure 3, from which it is clear that spherical configurations are not stable with their binding energies being generally negative. The binding energies for linear hydrogen self-clusters are of small positive values, suggesting that the self-clustering be energetically favorable. Specifically for the linear hydrogen cluster, hydrogen atoms prefer to arrange along $<$110$>$ directions. This arrangement ensures that all closest H-H pairs are in the most stable binding state. From Figure 3, we also observe that the planar structure appears to be the most stable structure for hydrogen self-clustering (except for n=3). Particularly worth noting is that the binding energy for planar self-cluster overall increases with the cluster size. This increasing tendency indicates that the ability of a $H_n$ platelet to capture isolated interstitial hydrogen atoms grows with its self-cluster size n, which may trigger a cascading procedure of hydrogen self-clustering at high hydrogen concentration environments. In the planar self-cluster the hydrogen atoms prefer to aggregate within $\{1 0 0\}$ planes, which ensures that all closest H-H pairs are in the most stable $<$110$>$ binding state, similar to the case of the linear hydrogen self-cluster. One additional thing worth noting is that, despite the $<$310$>$ H-H pairs being also energetically favorable, this pairing does not seem to show particular influence on the linear or planar aggregation of hydrogen atoms, likely owing to symmetry reasons.}
\subsection{Physical origin underlying H-H interaction}
 \begin{figure}[htp]
 \begin{center}
 \includegraphics[width=7cm]{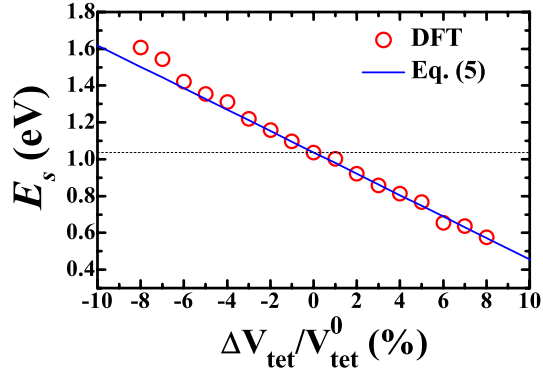}
 \end{center}
 \caption{(color online) Hydrogen solution energy as a function of tetrahedron volume change under hydrostatic strain.}
 \end{figure}

 The H-H interaction in metal mainly comprises two effects, i.e. the elastic interaction and the electronic interaction. {\color{black}The elastic interaction arises from the lattice dilatation induced by interstitial hydrogen, i.e., the introduction of an interstitial hydrogen into a perfect metal lattice distends its neighboring interstitial sites, which provides more space for subsequent insertion of interstitial hydrogen. Therefore, the combination of isolated hydrogen atoms into a hydrogen self-cluster can efficiently reduce the distortion in the lattice, consequently resulting in an attractive interaction among the interstitial hydrogen atoms. In order to assess the elastic interaction quantitatively, we first investigated the behavior of an interstitial hydrogen atom under hydrostatic strains. Figure 4 shows the hydrogen solution energy as a function of the strain. Interestingly, we find the solution energy varies linearly with the strain, and can be well described by\cite{Zhou2016}:
 \begin{equation}
 E_{s}=E_{s}^0-B\Omega_H\frac{\Delta V_{tet}}{V_{tet}^0}.
 \end{equation}
 Here, $E_{s}^0$ is hydrogen solution in tungsten when no strain is applied, $B$ is the bulk modulus, being 310 GPa for tungsten, $\Omega_H$ is the partial volume of a hydrogen atom in tungsten, being $\sim 3 \AA^3$ \cite{Zhou2016,Fukai-book}, $V_{tet}^0$ is the tetrahedron volume in the perfect tungsten lattice, and $\Delta V_{tet}$ is the tetrahedron volume change induced by the hydrostatic strain. This equation give us a simple analytical means to evaluate the elastic interaction between interstitial hydrogen atoms. According to our definition in Eqs. (1-2), we can estimate the elastic contribution of $E_b^{H_n}$ using an elastic binding energy:
 \begin{equation}
 E_{b}^{ELA}=B\Omega_H\frac{\Delta V_{tet}}{V_{tet}^0},
 \end{equation}
 where $\Delta V_{tet}$ is the volume change of the neighboring tetrahedron induced by a $H_{n-1}$ self-cluster.}

 The electronic interaction comes from the change of the hydrogen states in the electronic structure. For two hydrogen ions in close vicinity, the bonding and antibonding states between them will shift asymmetrically and raise the total energy. \cite{Becquart2009,Ouyang2011} Consequently, this leads to a repulsive interaction between H-H pairs. The electronic interaction can also be characterized by an electronic binding energy, $E_{b}^{ELE}$. Unfortunately, to the best of our knowledge, there is no analytical formula to evaluate the electronic binding energy. Here, the $E_{b}^{ELE}$ is estimated by the binding energies of hydrogen atoms in a frozen lattice where both supercell and atom positions are fixed.

 \begin{figure}[!htb]
 \begin{center}
 \includegraphics[width=14cm]{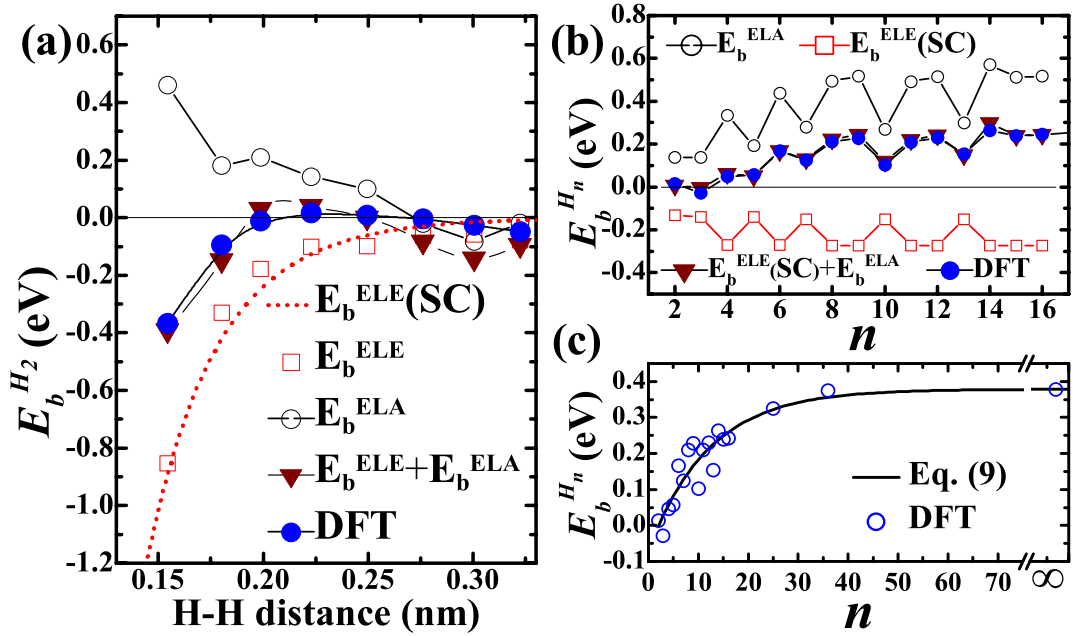}
 \end{center}
 \caption{(color online) (a) Elastic (ELA) and electronic (ELE) contributions of H-H binding energies at different H-H separation distance. SC represents electronic binding energy fitted using Eq. (7). (b) Binding energies of the most stable planar $H_{2\sim16}$ self-clusters, along with a similar analysis of the corresponding elastic and electronic contributions. (c) The evolution of binding energy of a $H_n$ cluster as the cluster size $n$varies, and the fitted curve (solid black line) using Eq. (9). For n=25 and 36, DFT results are calculated using a $8\times8\times4$ supercell with a $1\times1\times2$ \emph{k}-point grid. For $n\rightarrow\infty$, DFT result is calculated according to Eq. (8).}
 \end{figure}

 The binding energies, and corresponding elastic and electronic contributions associated with hydrogen self-clusters are shown in Figure 5. Let's start by examining the situation of H-H pairs, i.e., $H_2$ self-clusters, in tungsten (cf. Figure 5a). We can see that the sum of $E_{b}^{ELA}$ and $E_{b}^{ELE}$ is in excellent agreement with the DFT-calculated binding energy. This good agreement confirms that the H-H interaction in tungsten is indeed contributed by the elastic and electronic interactions. In particular the elastic interaction is generally positive while the electronic interaction is negative, which suggest that they respectively correspond to attraction and repulsion between two hydrogen atoms. Both elastic and electronic interactions decrease in magnitude as the H-H distance increases, with the electronic interaction being much stronger at short distance, but decaying more rapidly with the distance. The domination of electronic repulsion at short distances explains the fact that two interstitial hydrogen atoms are mutually exclusive in short distances ($<0.2$ nm). With the H-H distance increasing from 0.2 to 0.25 nm, the elastic attraction starts to surpass the electronic repulsion. Thus, the H-H pairs are in binding states at this range. With further increasing distance, both interactions approach zero, and the elastic interaction also changes to being negative, rendering the overall H-H interaction weak repulsion. Consequently, it can be concluded that the competition and interplay of the elastic and electronic contributions prescribe the behaviors of H-H pairs, and subsequently are expected to control the behaviors of hydrogen self-clusters. In this regard, below we perform a similar analysis for hydrogen self-clusters to elucidate the physical origin of hydrogen self-clustering.

 One particular interesting observation to note from Figure 4a is that the electronic interaction can be well fitted with a screened Coulomb (SC) potential\cite{Kittle-book}:
 \begin{equation}
 E_{b}^{ELE}(SC)=-\frac{q^2}{4\pi \varepsilon_0 r}\exp(-k_sr),
 \end{equation}
 where $q$ corresponds to the effective charge on each hydrogen, $\varepsilon_0$ is the vacuum permittivity, $r$ is the H-H separation distance, and $k_s=\sqrt{\frac{4}{a_0}(\frac{3n_0}{\pi})^{\frac{1}{3}}}$ is the screening paramete with $n_0$ being the electron density and $a_0$ being the Bohr radius \cite{Kittle-book}. Here, all $5d$ and $6s$ electrons in tungsten are treated as free electrons, which yield $k_s=231.6\ nm^{-1}$. Fitting the DFT-computed $E_b^{ELE}$ data to Eq. (7), we obtained $q=1.85\ e$, being very close to the hydrogen charge estimated by the Bader's charge analysis \cite{Henkelman2006,Sanville2007}, $q=1.86\ e$. This good agreement suggests that the electronic interaction between hydrogen atoms can be well estimated with the screened Coulomb interaction. Eq. (7) provides a simple analytical means to evaluate $E_{b}^{ELE}$, and is used below in the analysis of binding energies of hydrogen self-clusters.

 Figure 5b presents the binding energies of an additional interstitial hydrogen atom with the stable $H_{n-1}$ cluster. The corresponding elastic and electronic interactions respectively predicted from Eqs. (6) and (7) are also shown for comparison. The electronic interaction for a $H_n$ self-cluster is calculated by summing up all the pairwise $E_b^{ELE}$ terms (from Eq. (7)) between the $n^{th}$ hydrogen atom and each hydrogen atom in the $H_{n-1}$ cluster. Similar to what was previously observed in Figure 5a, the sum of elastic and electronic contributions well agrees with the DFT-calculated binding energy, again showing that the elastic and electronic interactions prescribe the H-H interaction in tungsten. Figure 5b also demonstrates that Eqs. (6) and (7) provides a description model to quantitatively analyze the interstitial hydrogen self-clustering in tungsten lattice.

 We note from Figure 4b that the electronic interaction only alternate between two distinct energy states of -0.14 eV and -0.27 eV in the range of self-cluster sizes investigated. {\color{black}This is because the electronic interaction is much localized and becomes negligible for H-H distance beyond 0.3 nm. Consequently the newly introduced hydrogen atom, i.e., the $n^{th}$ hydrogen, only interacts with its nearest-neighboring hydrogen atoms in the $H_{n-1}$ self-cluster. For instance, for the case shown in Figure 2c, in term of electronic interaction, the $16^{th}$ hydrogen only interacts with the $12^{th}$ and the $13^{th}$ hydrogen in the $H_{15}$ cluster, as the distance from any other hydrogen to the $16^{th}$ hydrogen is $>0.32$ nm. As the self-cluster size $n$ varies, there exist only two scenarios for the electronic interaction, being that the $n^{th}$ hydrogen interacts with one (for $n$=2, 3, 5, 7, 10, 13, ...) or two (for $n$=4, 6, 8, 9, 11, 12, 14, 15, ...) nearest-neighboring hydrogen in the $H_{n-1}$ self-cluster. These two scenarios correspond to the two distinct energy states of -0.14 eV and -0.27 eV respectively, leading to the electronic interaction alternation as the $H_{n}$ self-cluster grows.}

 On the other hand, the elastic interaction overall exhibits an increasing trend as the self-cluster size $n$ increases. The trend indicates that the elastic attraction would gradually overcome the local electronic repulsion as the self-cluster grows. As we previously discussed, planar structure is the most favorable structure of hydrogen self-clusters, and the elastic interaction comes from the expansion of interstitial sites nearby a hydrogen cluster. {\color{black}Clearly, the insertion of a hydrogen platelet into tungsten lattice will cause sizable lattice expansion, particularly around the platelet edge. The appreciable expansion at the edges provides more space for interstitial hydrogen atoms, making it energetically more favorable for hydrogen occupancy, thus resulting in the planar growth of the hydrogen cluster. As the self-cluster grows, the degree of lattice swelling is expected to increase. However, presumably the lattice swelling would eventually plateau and it is anticipated that the binding energy of hydrogen will converge to a steady state, i.e., a constant value when the self-cluster size grows beyond a critical value. However, the determination of this critical value would be computationally time consuming. Nonetheless, we can evaluate the steady state by considering a self-cluster with an infinite number of hydrogen atoms, i.e., the tungsten lattice is divided into two parts by a mono-layer of hydrogen atoms ({\color{black}cf. Figure 2e}, this monolayer is referred to as $H_\infty$ in the equation below), and calculated the converged binding energy $E_b^{H_\infty}$ (i.e., $E_b^{H_n}$ with $n\rightarrow\infty$ ) by:
 \begin{equation}
 E_{b}^{H_\infty}=\frac{(N-1) E_{tot}^{bulk} + E_{tot}^{H_\infty} - NE^{H_1}_{tot}}{N}.
 \end{equation}
 In the above expression, $ E_{tot}^{bulk}$ and $E^{H_1}_{tot}$ are quantities previously defined in Eq. (1), $E_{tot}^{H_\infty}$ is the total energy of the super-cell that contains a mono-layer of hydrogen (i.e., $H_\infty$) that divides the tungsten lattice, and $N$ is the number of hydrogen atoms within $H_\infty$ in the super-cell. $E_{b}^{H_\infty}$ represents an average binding energy that can be considered as the steady-state binding energy of hydrogen for self-clusters that are beyond the critical size. According to our DFT calculations, this steady-state binding energy is found to be 0.38 eV when $n\rightarrow\infty$. The DFT-calculated data for different $n$ are shown in Figure 5. It is found that the data can be well fitted by a simple function, assuming the convergence scale as $\alpha \exp(\beta n)$ (cf. Figure 5c):
 \begin{equation}
 E_{b}^{H_n}=E_{b}^{H_\infty}-\alpha \exp(-n/\beta),
 \end{equation}
where the parameters $\alpha$ and $\beta$ are fitted to be 0.45 eV and 12.04, respectively. As we can see, a convergence in the binding energy is established as $n$ reaches around 30.}
\subsection{Thermodynamic and kinetic analyses of H self-clustering}
The above results demonstrate that hydrogen self-clustering is energetically possible in tungsten. Nevertheless, a comprehensive evaluation of the feasibility of H self-clustering under different temperatures and hydrogen concentrations necessitates further thermodynamic investigation, as elaborated below. During the formation of a $H_n$ self-cluster, the Gibbs free energy change of the system, denoted as $\Delta G_{n}$, is given as:
 \begin{equation}
 \Delta G_{n}=\Delta H -T\Delta S,
 \end{equation}
 where $\Delta H$ and $\Delta S$ are the enthalpy and entropy changes respectively associated with the H aggregation reaction, i.e., $H_1+H_1+H_1 \cdots \longrightarrow H_n$. According to our definition in Eq. (1), the enthalpy change can be evaluated as:
 \begin{equation}
 \Delta H=-\sum_{i=2}^n E_{b}^{H_i}.
 \end{equation}
 While the entropy change can be approximately given by\cite{Fukai-book}:
 \begin{equation}
 \Delta S=(n-1)k_{B}\ln\frac{C_H}{1-C_H},
 \end{equation}
 where $k_B$ is the Boltzmann constant, $C_H$ is bulk atomic concentration of hydrogen in tungsten (i.e., H/W ratio).

 Based on Eqs. (9-12), the Gibbs free energy change of the system as a function of the self-cluster size $n$ at different $C_H$ and $T=300$ K is plotted in Figure 6a. We see that the enthalpy term $\Delta H$ decreases monotonically as $n$ increases, with the rate of change in $\Delta H$ first accelerating and then gradually stabilized at a {\color{black}constant rate. On the other hand, the term $-T\Delta S$ increases almost linearly with a rate depending on the hydrogen concentration. As illustrated in Figure 6, the descending enthalpy as $n$ increases provides a driving force for hydrogen self-clustering, which competes with the entropy effect that favors dissociation. Consequently a critical hydrogen concentration $C_H^*$ exists, at which the enthalpy and entropy effects cancel each other out, and $\Delta G_{n}$ becomes a constant for large $H_n$ self-clusters. According to Eqs. (10-12), the critical concentration $C_H^*$ is given by:
 \begin{equation}
 C_H^*=\frac{1}{\exp(-E_{b}^{H_\infty}/k_{B}T)+1}.
 \end{equation}
 For $C_H<C_H^*$, $\Delta G_{n}$ increases monotonically as the self-cluster grows. On the other hand, for $C_H>C_H^*$, $\Delta G_{n}$ first increases and then decreases as the self-cluster grows, indicating that it is thermodynamically possible for hydrogen cluster nucleation. The value of $n$ where $\Delta G_{n}$ is maximized then corresponds to the critical nucleus size, denoted as $n_{nucl}$, and the value of $\Delta G_{n}$ at $n_{nucl}$ represents the associated nucleation barrier, marked by $\Delta G_{nucl}$.}

 \begin{figure}[!htb]
 \begin{center}
 \includegraphics[width=14cm]{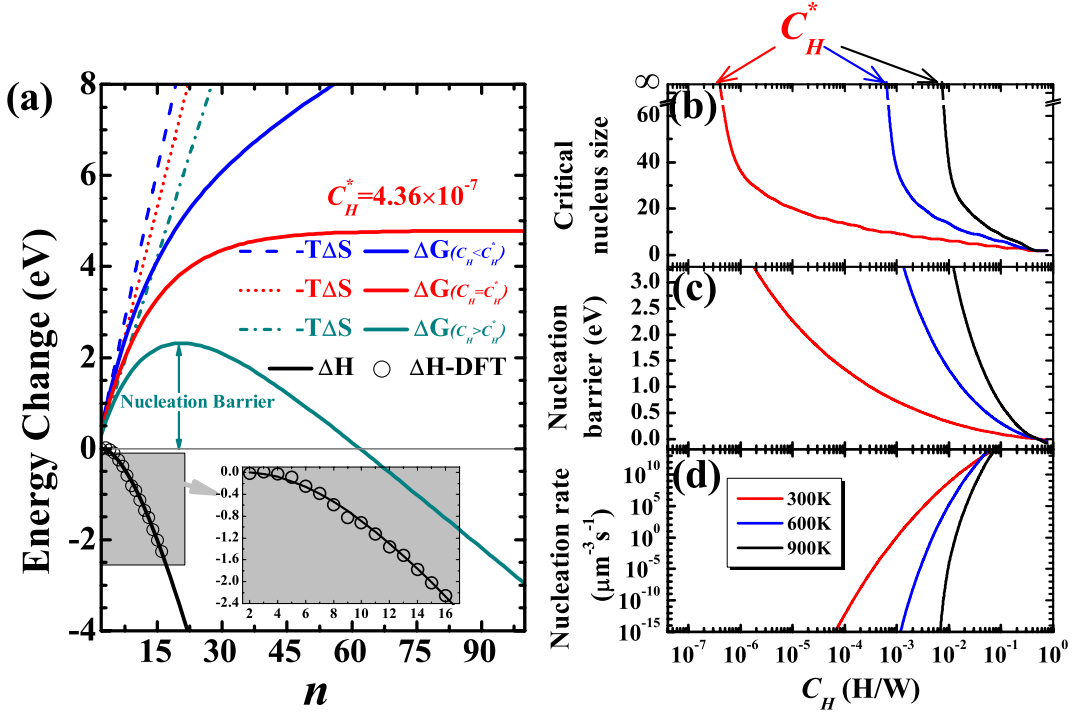}
 \end{center}
 \caption{(color online) (a) Changes of free energy, enthalpy, and entropy term during self-clustering of $H_n$ at 300K with different hydrogen concentrations, calculated according to Eqs. (9-12). Circles and lines represent DFT and analytical data, respectively. The red lines highlight the critical concentration at which the enthalpy and entropy cancels each other out. (b), (c), and (d) present the critical nucleus size, nucleation barriers, and nucleation rate of the hydrogen self-cluster as a function of hydrogen concentrations at temperature 300 K, 600 K, and 900 K, respectively.}
 \end{figure}

 Figures 6b and 6c respectively present $n_{nucl}$ and $\Delta G_{nucl}$ as functions of $C_H$ at different temperatures. As shown in Figure 6b, the critical concentration $C_H^*$ increases rapidly with the temperature, being $10^{-7}$ at 300 K and reaches to $10^{-3}$ at 600 K. These concentrations are far greater than the H solubility in tungsten ($10^{-23}$ at 300 K and $10^{-13}$ at 600 K under 1 bar of $H_2$, extrapolated according to Ref.\cite{Frauenfelder1969}). This suggests that the H self-clustering can not occur under normal conditions. The formation of hydrogen self-cluster becomes more difficult at high temperature, which is well expected because the entropy effect is stronger at higher temperatures. In addition, it is also clear from Figure 6 that the critical nucleus size and nucleation barrier drops rapidly with the increase of H concentration. The hydrogen self-cluster can accordingly be formed more easily as $C_H$ raises. We have further calculated the homogeneous nucleation rate of hydrogen self-clusters based on the classic nucleation theory \cite{Sear2007}:
 \begin{equation}
 \Gamma_{nucl}=(\rho C_H)^2 D_{H} R_{H} Z\exp (\frac{-\Delta G_{nucl}}{k_BT}),
 \end{equation}
 where $\rho$ is the atomic density of tungsten, $D_{H}=5.7\times10^{-8}\exp(-0.22eV/k_{B}T)\ m^2/s$ is the diffusivity of hydrogen in tungsten,\cite{Kong2015-2} and $R_{H}=0.223$ nm is the interaction range of $H_1$. Z denotes the Zeldovich factor, being approximately a constant, $Z=0.1$.\cite{Sear2007} The nucleation rates at different temperatures as functions of $C_H$ are shown in Figure 6d, where we can see that the nucleation rate of the hydrogen self-cluster increases rapidly with $C_H$. Therefore, in higher hydrogen concentration environments, the hydrogen self-clustering is not only thermodynamically more favorable but also kinetically more feasible.
\subsection{Possible mechanism of hydrogen bubble nucleation via self-clustering }
 From the above, we have shown that a necessary condition for $H_n$ self-cluster formation is to have hydrogen concentration reach the critical concentration $C_H^*$. Though it is normally difficult for the hydrogen concentration to reach $C_H^*$ due to the extremely low hydrogen solubility in tungsten, it can be possible when tungsten is exposed to low-energy high flux hydrogen plasma when the hydrogen concentration can be significantly enhanced. Generally, the maximum concentration of hydrogen at the material surface during ion implantation, $C_{H}^{Ion}$, can be evaluated by\cite{Kolasinski2015}:
 \begin{equation}
 C_{H}^{Ion}=Xf/D_{H},
 \end{equation}
 where $f$ is the incident flux of hydrogen ion. Based upon our calculations using the TRIM code\cite{Ziegler1985}, the implantation depth $X$ is approximately proportional to the incident ion energy, i.e., $X=E_{H}^{ion}\times 0.03\ nm/eV$.

 \begin{figure}[!htb]
 \begin{center}
 \includegraphics[width=7cm]{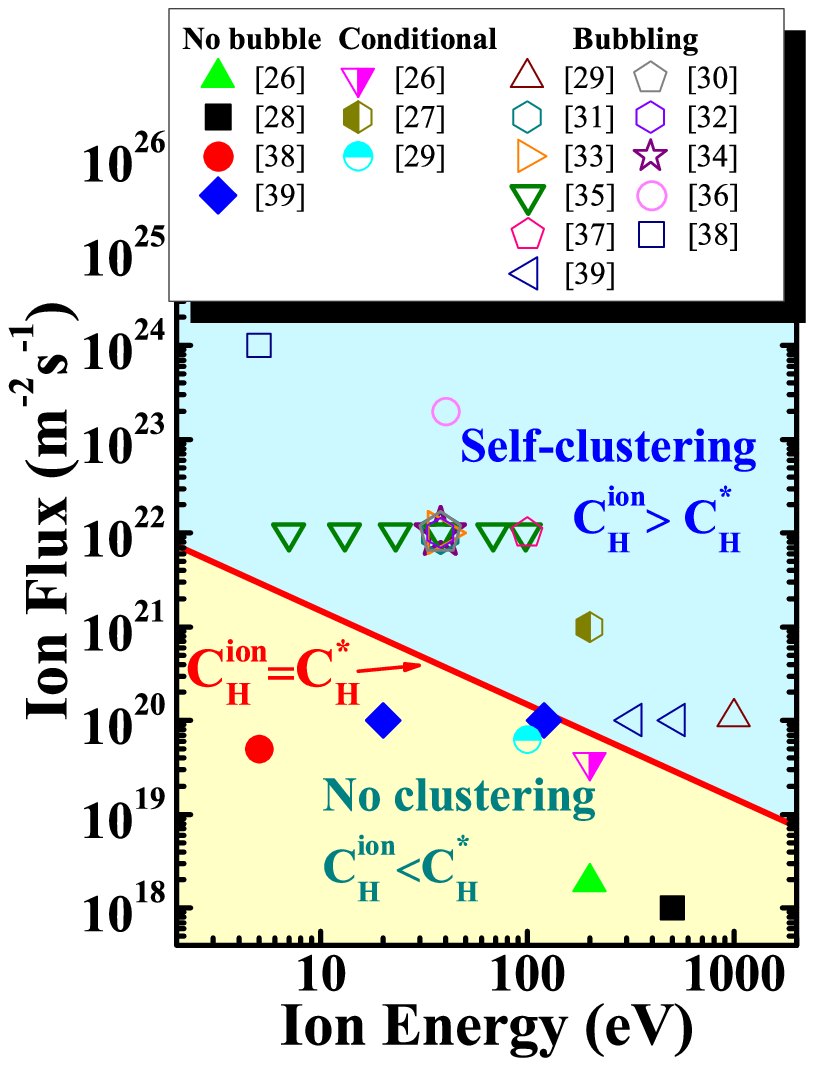}
 \end{center}
 \caption{(color online) H self-clustering ability under different H plasma loading fluxes and energies, calculated according to Eqs. (13) and (15). The cyan colored region maps the conditions that can result in self-clustering while the yellow colored region corresponds to no-clustering conditions. The symbols denote experimental data at $\sim300$ K from Ref.\cite{Alimov2005,Alimov2008,Poon2002,wang2001,Shu2007-2,Shu2007-1,Shu2008,Alimov2009,Alimov2012,Luo2005,Zibrov2017,Tokunaga2005,Zayachuk2015,Ni2015}. Hollow ones represent bubble formation, filled ones represent no evident bubble formation, half-filled ones represent that bubble formation also depends on variables other than ion energy and flux.}
 \end{figure}

In this way, the plasma flux and energy required for hydrogen self-clustering can be quantitatively predicted, i.e., $C_{H}^{Ion}>C_H^*$. Figure 7 shows the results of our prediction at the room temperature, in comparison with available experimental results \cite{Alimov2005,Alimov2008,Poon2002,wang2001,Shu2007-2,Shu2007-1,Shu2008,Alimov2009,Alimov2012,Luo2005,Zibrov2017,Tokunaga2005,Zayachuk2015,Ni2015} (pre-annealed undamaged tungsten subjected to low energy pure hydrogen plasma at around room temperature). The plot is divided into two regions by the condition of $C_{H}^{Ion}=C_H^*$, respectively corresponding to where hydrogen self-clustering is predicted to occur (i.e., 'self-clustering' region) and not to occur ('no clustering' region). As seen in Figure 7, the loading experimental plasma conditions that cause significant bubble formation (indicated by hollow symbols) fall well within the self-clustering region, while those without evident bubble observation (indicated by solid symbols) fall within the no-clustering region. This excellent agreement between experimental observations and theoretical prediction suggests that hydrogen self-clustering be closely related to hydrogen bubble formation.

 \begin{figure}[!htb]
 \begin{center}
 \includegraphics[width=14cm]{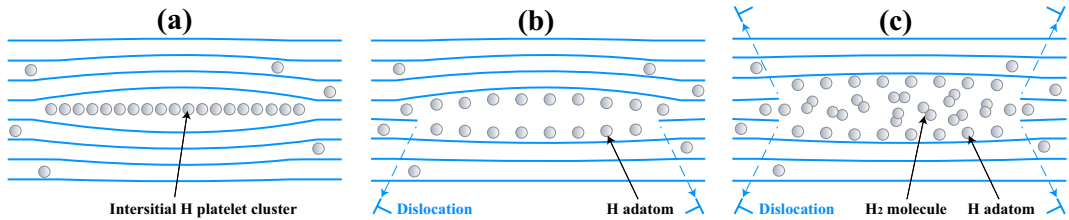}
 \end{center}
 \caption{(color online)  Schematic representation of the process of platelet aggregation of hydrogen atoms and spontaneous bubble formation in the perfect tungsten lattice. The hydrogen atoms are represented by small white balls.}
 \end{figure}

 {\color{black}The connection between hydrogen self-clustering and hydrogen bubble formation was also reflected by a few earlier experiments \cite{Tetelman1963,Marrow1996,Chen1990}, where cleavage cracks in bcc iron were induced by electrochemical hydrogen charging even when no external strain was applied. The cleavage cracks were observed to initiate and propagate along $\{001\}$$<$110$>$ directions, accompanied by clear dislocation emission. In light of these experiments and our results, below we propose a possible mechanism for the hydrogen bubble nucleation by hydrogen self-clustering.} When excessive interstitia hydrogen atoms exist in the tungsten lattice, they would gather together to form thin platelets along the $\{001\}$ lattice planes to reduce the strain energy of the system. Once the platelet hydrogen self-cluster formed, it would induce noticeable lattice bending and weaken metallic bonds around the platelet (see Figure 8a). As suggested by Fujita,\cite{Fujita1976} dislocations can be created at the platelet edge so as to further release the elastic energy. Meanwhile, this would make room for interstitial hydrogen atoms in the platelet self-cluster to convert to hydrogen adatoms (see Figure 8b), which can lower the energy. This would further increase the lattice bending around the platelets, induce more severe loosening of the lattice cohesion, eventually causing a slight opening of the lattice. As the above process continues, adatoms would accumulate and get desorbed to produce $H_2$ gas in the opening. The internal gas pressure would conduce to the further growth of the opening and ultimately give rise to spontaneous crack (hydrogen bubble) formation (see Figure 8c). %{\color{black}This mechanism allows cracks or bubbles to nucleate in a defect-free lattice, which provides an quantitative explanation for the low energy hydrogen plasma induced bubble formation in annealed pure tungsten.}
\section{Conclusions}
In summary, this work studied the self-clustering of hydrogen atoms within the tungsten lattice using first-principles calculations. Possible hydrogen self-cluster structures have been examined and it was found that hydrogen atoms would form two-dimensional platelet-like structures, preferentially along $\{100\}$ planes. The hydrogen self-cluster was shown to be energetically more favorable as the cluster size increases, suggesting the increasing ability of a hydrogen self-cluster to capture additional interstitial hydrogen as the cluster grows. These hydrogen self-clustering behaviors can be well understood by the competition between long-ranged elastic attraction and local electronic repulsion, {\color{black}which were both isolated and evaluated quantitatively.} Further thermodynamic analysis shows that there exists a critical hydrogen concentration, above which hydrogen self-clusters are thermodynamically stable and kinetically feasible, {\color{black}and below which all hydrogen self-clusters become unfavorable.} Based on this critical hydrogen concentration, the plasma loading conditions under which hydrogen self-clusters form are predicted. Our theoretical predictions show excellent agreement with experimental results of hydrogen bubble formation in tungsten exposed to low-energy hydrogen irradiation, suggesting close connections between hydrogen self-clustering and hydrogen bubble formation. Based on our findings, {\color{black}we proposed that the lattice bending induced by platelet-like hydrogen self-clusters may triggers dislocation emissions to release the elastic energy, which provides rooms for $H_2$ gas precipitation, and finally leads to spontaneous bubble formation.} The present study provides mechanistic insights and quantitative models towards understanding of plasma-induced hydrogen bubble formation in plasma-facing metals.
\section*{Acknowledgement}
This work is supported by the National Key Research and Development Program of China (Grant No.: 2017YFA0402800), the National Natural Science Foundation of China (Nos.: 11505229, 11735015), and by the Natural Sciences and Engineering Research Council of Canada (NSERC) Discovery grant (Grant No. RGPIN 418469-2012). We also thank the computing power from the Center for Computation Science, Hefei Institutes of Physical Sciences, and the Supercomputer Consortium Laval UQAM McGill and Eastern Quebec. J. Hou. acknowledges the financial support from China Scholarship Council (CSC).

\section*{References}

\renewcommand*{\bibfont}{\footnotesize}
\setlength{\bibsep}{0pt plus 0.3ex}
\bibliography{unsrt}

\begin{thebibliography}{00}
\bibitem{Xie2015}D.G. Xie, Z.J. Wang, J. Sun, J. Li, E. Ma, Z.W. Shan, In situ study of the initiation of hydrogen bubbles at the aluminium metal/oxide interface, Nat. Mater., 14 (2015) 899-903.
\bibitem{Tiegel2016}M.C. Tiegel, M.L. Martin, A.K. Lehmberg, M. Deutges, C. Borchers, R. Kirchheim, Crack and blister initiation and growth in purified iron due to
hydrogen loading, Acta Mater., 115 (2016) 24-34.
%\bibitem{Yen2003}S.K. Yen, I.B. Huang, Mater. Chem. Phys., 80 (2003) 662-666
%\bibitem{Panagopoulos1998}C.N. Panagopoulos, A.S. El-Amoush, P.E. Agathocleous, Corros. Sci., 40 (1998) 1837-1844
\bibitem{Rozenak2007}P. Rozenak, Hemispherical bubbles growth on electrochemically charged aluminum with hydrogen, Int. J. Hydrog. Energy 32 (2007) 2816-2823.
\bibitem{Condon1993}J.B. Condon, T. Schober, Hydrogen bubbles in metals, J. Nucl. Mater., 207 (1993) l-24.
\bibitem{Becquart2006}C.S. Becquart, C. Domain, Migration Energy of He in W Revisited by Ab Initio Calculations, Phys. Rev. Lett., 97 (2006) 196402.
\bibitem{Kong2016-2}X.S. Kong, Y. You, X. Li, X. Wu, C.S. Liu, J.L. Chen, G.N. Luo, Towards understanding the differences in irradiation effects of He, Ne and Ar plasma by investigating the physical origin of their clustering in tungsten, Nucl. Fusion, 56 (2016) 106002.
\bibitem{Kong2016-1}X.S Kong, X. Wu, C.S. Liu, Q. F. Fang, Q.M. Hu, J.L. Chen, G.N. Luo, First-principles calculations of transition metal solute interactions with hydrogen in tungsten, Nucl. Fusion, 56 (2016) 026004.
\bibitem{Henriksson2005}K. O. E. Henriksson, K. Nordlund, A. Krasheninnikov, J. Keinonen, Difference in formation of hydrogen and helium clusters in tungsten, Appl. Phys. Lett., 87 (2005) 163113.
\bibitem{Becquart2009}C.S. Becquart, C. Domain, A density functional theory assessment of the clustering behaviour of He and H in tungsten, J. Nucl. Mater., 386-388 (2009) 109-111.
\bibitem{Liu2009}Y.L. Liu, Y. Zhang, G.N. Luo, G.H. Lu, Structure, stability and diffusion of hydrogen in tungsten: A first-principles study, J. Nucl. Mater., 390-391 (2009) 1032-1034.
\bibitem{You2014}Y.W. You, D. Li, X.S. Kong, X. Wu, C.S. Liu, Q.F. Fang, B.C. Pan, J.L. Chen, G.N. Luo, Clustering of H and He, and their effects on vacancy evolution in tungsten in a fusion environment, Nucl. Fusion, 54 (2014) 103007.
\bibitem{Monasterio2009}P.R. Monasterio, T.T. Lau, . Yip, K.J. Van Vliet, Hydrogen-Vacancy Interactions in Fe-C Alloys, Phys. Rev. Lett., 103 (2009) 085501.
\bibitem{Ouyang2011}C. Ouyang, Y.S. Lee, Hydrogen-induced interactions in vanadium from first-principles calculations, Phys. Rev. B, 83 (2011) 045111.
\bibitem{Pezold2011}J. von Pezold, L. Lymperakis, J. Neugebeauer, Hydrogen-enhanced local plasticity at dilute bulk H concentrations: The role of H-H interactions and the formation of local hydrides, Acta Mater., 59 (2011) 2969-2980.
\bibitem{Zhou2016}X. Zhou, D. Marchand, D.L. McDowell, T. Zhu, J. Song, Chemomechanical Origin of Hydrogen Trapping at Grain Boundaries in fcc Metals, Phys. Rev. Lett., 116 (2016) 075502.
\bibitem{Valles2017}G. Valles, M. Panizo-Laiz, C. Gonzalez, I. Martin-Bragado, R. Gonzalez-Arrabal, N. Gordillo, R. Iglesias, C.L. Guerrero, J.M. Perlado, A. Rivera, Influence of grain boundaries on the radiation-induced defects and hydrogen in nanostructured and coarse-grained tungsten, Acta Mater., 122 (2017) 277-286.
\bibitem{Zhou2010}H.B. Zhou, Y.L. Liu, S. Jin, Y. Zhang, G.N. Luo, G.H. Lu, Investigating behaviours of hydrogen in a tungsten grain boundary by first principles: from dissolution and diffusion to a trapping mechanism, Nucl. Fusion, 50 (2010) 025016.
\bibitem{Xiao2012}W. Xiao, W.T. Geng, Role of grain boundary and dislocation loop in H blistering in W: A density functional theory assessment, J. Nucl. Mater., 430 (2012) 132-136.
\bibitem{Kong2013}X.S. Kong, Y.W. You, Q.F. Fang, C.S. Liu, J.L. Chen, G.N. Luo, B.C. Pan, Z. Wang, The role of impurity oxygen in hydrogen bubble nucleation in tungsten, J. Nucl. Mater., 433 (2013) 357-363.
\bibitem{Geng2017}W.T. Geng, L. Wan, J.P. Du, A. Ishii, N. Ishikawa, H. Kimizuka, S. Ogata, Hydrogen bubble nucleation in $\alpha$-iron, Scr. Mater., 134 (2017) 105-109.
\bibitem{Sun2016}L. Sun, S. Jin, G.H. Lu, L. Wang, High hydrogen retention in the sub-surfaces of tungsten plasma facing materials: A theoretical insight, Scr. Mater., 122 (2016) 14-17.
\bibitem{Tetelman1963}A.S. Tetelman, W.D. Robertson, Direct observation and analysis of crack propagation in Iron-3$\%$ silicon single crystals, Acta Metall., 11 (1963) 415-426.
\bibitem{Ren2008}X.C. Ren, Q.J. Zhou, G.B. Shan, W.Y. Chu, J.X. Li, Y.J. Su, L.J. Qiao, A Nucleation Mechanism of Hydrogen Blister in Metals and Alloys, Metall. Mater. Trans. A, 39 (2008) 87-97.
\bibitem{Escobar2011}D. P. Escobar, C. Minambres, L. Duprez, K. Verbeken, M. Verhaege, Internal and surface damage of multiphase steels and pure iron after electrochemical hydrogen charging, Corrosion Science 53 (2011) 3166-3176.
\bibitem{Poon2005}M. Poon, R.G. Macaulay-Newcombe, J.W. Davis, A.A. Haasz, Effects of background gas impurities during D$^+$ irradiation on D trapping in single crystal tungsten, J. Nucl. Mater., 337-339 (2005) 629-633.
\bibitem{Alimov2005} V.Kh. Alimov, J. Roth, M. Mayer, Depth distribution of deuterium in single and polycrystalline tungsten up to depths of several micrometers, J. Nucl. Mater., 337-339 (2005) 619-623.
\bibitem{Alimov2008} V.Kh. Alimov, J. Roth, R.A. Causey, D.A. Komarov, Ch. Linsmeier, A. Wiltner, F. Kost, S. Lindig, Deuterium retention in tungsten exposed to low-energy, high-flux
clean and carbon-seeded deuterium plasmas, J. Nucl. Mater., 375 (2008) 192-201.
\bibitem{Poon2002} M. Poon, R.G. Macaulay-Newcombe, J.W. Davis, A.A. Haasz, Flux dependence of deuterium retention in single crystal tungsten, J. Nucl. Mater., 307-311 (2002) 723-728.
\bibitem{wang2001} W. Wang, J. Roth, S. Lindig, C.H. Wu, Blister formation of tungsten due to ion bombardment, J. Nucl. Mater., 299 (2001), 124-131.
\bibitem{Shu2007-2}W.M. Shu, E. Wakai, T. Yamanishi, Blister bursting and deuterium bursting release from tungsten exposed to high fluences of high flux and low energy deuterium plasma, Nucl. Fusion, 47 (2007) 201-209.
\bibitem{Shu2007-1} W.M. Shu, A. Kawasuso, Y. Miwa, E. Wakai, G.N. Luo, T. Yamanishi,  Microstructure dependence of deuterium retention and blistering in the near-surface region of tungsten exposed to high flux deuterium plasmas of 38 eV at 315K, Phys. Scr., T128 (2007) 96-99.
\bibitem{Shu2008} W.M. Shu, K. Isobe, T. Yamanishi, Temperature dependence of blistering and deuterium retention in tungsten exposed to high-flux and low-energy deuterium plasma, Fusion Eng. Des., 83 (2008) 1044-1048.
\bibitem{Alimov2009} V.Kh. Alimov, W.M. Shu, J. Roth, K. Sugiyama, S. Lindig, M. Balden, K. Isobe, T. Yamanishi, Surface morphology and deuterium retention in tungsten exposed to low-energy, high flux pure and helium-seeded deuterium plasmas, Phys. Scr., T138 (2009) 014048.
\bibitem{Alimov2012} V.Kh. Alimov, B. Tyburska-P¨¹sche, S. Lindig, Y. Hatano, M. Balden, J. Roth, K. Isobe, M. Matsuyama, T. Yamanishi, Temperature dependence of surface morphology and deuterium retention in polycrystalline ITER-grade tungsten exposed to low-energy, high-flux D plasma. J. Nucl. Mater.,420 (2012) 519-524.
\bibitem{Luo2005} G.N. Luo, W.M. Shu, M. Nishi, Incident energy dependence of blistering at tungsten irradiated by low energy high flux deuterium plasma beams, J. Nucl. Mater., 347 (2005) 111-117.
%\bibitem{Ni2015} W. Ni, Q. Yang, H. Fan a, L. Liu, T. Berthold, G. Benstetter, D.Liu, J. Nucl. Mater., 464 (2015) 216-220
\bibitem{Zibrov2017}M. Zibrov, M. Balden, T.W. Morgan, M. Mayer, Deuterium trapping and surface modification of polycrystalline tungsten exposed to a high-flux plasma at high fluences, Nucl. Fusion, 57 (2017) 046004.
\bibitem{Tokunaga2005}K. Tokunaga, M.J. Baldwin, R.P. Doerner, N. Noda, Y. Kubota, N. Yoshida, T. Sogabe, T. Kato, B. Schedler, Blister formation and deuterium retention on tungsten exposed to low energy and high flux deuterium plasma, J. Nucl. Mater., 337-339 (2005) 887-891.
\bibitem{Zayachuk2015} Y. Zayachuk, A. Manhard, M.H.J. ¡®t Hoen, W. Jacob, P.A. Zeijlmans van Emmichoven, G. Van Oost, The effect of ion flux on plasma-induced modification and deuterium retention in tungsten and tungsten-tantalum alloys, J. Nucl. Mater., 464 (2015) 69-72.
%\bibitem{Alimov2005} V.Kh. Alimov, J. Roth, M. Mayer, J. Nucl. Mater., 337-339 (2005) 619-623
%\bibitem{Alimov2008} V.Kh. Alimov, J. Roth, R.A. Causey, D.A. Komarov, Ch. Linsmeier, A. Wiltner, F. Kost, S. Lindig, J. Nucl. Mater., 375 (2008) 192-201
%\bibitem{Poon2002} M. Poon, R.G. Macaulay-Newcombe, J.W. Davis, A.A. Haasz, J. Nucl. Mater., 307-311 (2002) 723-728
%\bibitem{Poon2005}M. Poon, R.G. Macaulay-Newcombe, J.W. Davis, A.A. Haasz, J. Nucl. Mater., 337-339 (2005) 629-633
%\bibitem{wang2001} W. Wang, J. Roth, S. Lindig, C.H. Wu, J. Nucl. Mater., 299 (2001), 124-131
\bibitem{Ni2015} W. Ni, Q. Yang, H. Fan a, L. Liu, T. Berthold, G. Benstetter, D.Liu, Ordered arrangement of irradiation-induced defects of polycrystalline tungsten irradiated with low-energy hydrogen ions, J. Nucl. Mater., 464 (2015) 216-220.
\bibitem{Buzi2014}L. Buzi, G. De Temmerman, B. Unterberg, M. Reinhart, A. Litnovsky, V. Philipps, G. Van Oost, S. Moller, Influence of particle flux density and temperature on surface modifications of tungsten and deuterium retention, J. Nucl. Mater., 455 (2014) 316-319.
\bibitem{Buzi2015}L. Buzi, G. De Temmerman, B. Unterberg, M. Reinhart, T. Dittmar, D. Matveev, Ch. Linsmeier, U. Breuer, A. Kreter, G. Van Oost, Influence of tungsten microstructure and ion flux on deuterium plasma-induced surface modifications and deuterium retention, J. Nucl. Mater., 463 (2015) 320-324.
\bibitem{Jia2017}Y.Z. Jia, W. Liu, B. Xu, S.L. Qu, L.Q. Shi, T.W. Morgan, Subsurface deuterium bubble formation in W due to low-energy high flux deuterium plasma exposure, Nucl. Fusion, 57 (2017) 034003.
%\bibitem{Roth2009}J. Roth, E. Tsitrone, A. Loarte, et al., J. Nucl. Mater., 390-391 (2009) 1-9
%\bibitem{Poon2005}M. Poon, R.G. Macaulay-Newcombe, J.W. Davis, A.A. Haasz, J. Nucl. Mater., 337-339 (2005) 629-633
\bibitem{Fukai2003}Y. Fukai, Formation of superabundant vacancies in M-H alloys and some of its consequences: a review, J. Alloys Compd., 356-357 (2003) 263-269.
\bibitem{Sugimoto2014}H. Sugimoto, Y. Fukai, Hydrogen-induced superabundant vacancy formation in bcc Fe: Monte Carlo simulation, Acta Mater., 67 (2014) 418-429.
\bibitem{Sugimoto2017}H. Sugimoto, Y. Fukai, Hydrogen-induced superabundant vacancy formation by electrochemical methods in bcc Fe: Monte Carlo simulation, Scr. Mater., 134 (2017) 20-23.
%\bibitem{Kato2015}D. Kato, H. Iwakiri, Y. Watanabe, K. Morishita, T. Muroga, Nucl. Fusion 55 (2015) 083019
%\bibitem{Fernandez2015}N. Fernandez, Y. Ferro, D. Kato, Acta Mater., 94 (2015) 307-318
%\bibitem{Scamans1979}G. M. Scamans, A. S. Rehal, J. Mater. Sci. 14 (1979) 2459-2470.
\bibitem{Fujita1976}F.E. Fujita, the role of Hydrogen in the fracture of iron and steel, Trans. JIM., 17 (1976) 232-238.
\bibitem{Kamachi1972} K. Kamachi, S.Miyata, X-ray study of hydrogen-induced phenomena affecting mechanical behaviors of austenitic stainless steels, Proc. Int. Conf. Mech. Behaviors of Materials, 3 (1972) 274.
\bibitem{Kresse1993}G. Kresse, J. Hafner, Ab initio molecular dynamics for liquid metals, Phys. Rev. B, 47 (1993) 558-561.
\bibitem{Kresse1996}G. Kresse, J. Furthmuller, Efficient iterative schemes for ab initio total-energy calculations using a plane-wave basis set, Phys. Rev. B, 54 (1996) 11169-11186.
\bibitem{Blochl1994}P.E. Blochl, Projector augmented-wave method, Phys. Rev. B 50 (1994) 17953.
\bibitem{Perdew1992}J.P. Perdew, J.A. Chevary, S.H. Vosko, K.A. Jackson, M.R. Pederson, D.J. Singh, Carlos Fiolhais, Atoms, molecules, solids, and surfaces: Applications of the generalized gradient approximation for exchange and correlation, Phys. Rev. B 46 (1992) 6671.
\bibitem{Perdew1993}J.P. Perdew, J.A. Chevary, S.H. Vosko, K.A. Jackson, M.R. Pederson, D.J. Singh, Carlos Fiolhais, Erratum: Atoms, molecules, solids, and surfaces: Applications of the generalized gradient approximation for exchange and correlation, Phys. Rev. B 48 (1993) 4978(E).
\bibitem{Switendick1979}A. C. Switendick, Band Structure Calculations for Metal Hydrogen Systems, Zeitschrift f¨¹r Physikalische Chemie 117 (1979) 89.
\bibitem{Fukai-book}Y. Fukai,The Metal-Hydrogen System-Basic Bulk Properties, Second, Revised and Updated Edition, Springer, 2005.
\bibitem{Kittle-book} C. Kittle, Introduction to Solid State Physics, 6th Ed., Wiley, New York, 1986.
\bibitem{Henkelman2006} G. Henkelman, A. Arnaldsson, H. Jonsson, A fast and robust algorithm for Bader decomposition of charge density, Comput. Mater. Sci. 36 (2006) 354-360.
\bibitem{Sanville2007}E. Sanville, S.D. Kenny, R. Smith, G. Henkelman, Improved grid-based algorithm for Bader charge allocation, J. Comput. Chem. 28 (2007) 899.
\bibitem{Frauenfelder1969} R. Frauenfelder, Solution and Diffusion of Hydrogen in Tungsten, J. Vac. Sci. Technol., 6 (1969) 388.
\bibitem{Sear2007}R. P. Sear, Nucleation: theory and applications to protein solutions and colloidal suspensions, J. Phys.: Condens. Matter, 19 (2007) 033101.
\bibitem{Kong2015-2}X.S. Kong, S. Wang, X. Wu, Y.W. You, C.S. Liu, Q.F. Fang, J.L. Chen, G.N. Luo, First-principles calculations of hydrogen solution and diffusion in tungsten: Temperature and defect-trapping effects, Acta Mater., 84 (2015) 426-435.
\bibitem{Kolasinski2015}R. D. Kolasinski, M. Shimada, Y. Oya, D. A. Buchenauer, T. Chikada, D. F. Cowgill, D. C. Donovan, R. W. Friddle, K. Michibayashi, M. Sato, A multi-technique analysis of deuterium trapping and near-surface precipitate growth in plasma-exposed tungsten, J. Appl. Phys., 118 (2015) 073301.
\bibitem{Ziegler1985}J. F. Ziegler, J. P. Biersack, U. Littmark, The Stopping and Range of Ions in Solids, Pergamon Press, New York, 1985.
\bibitem{Marrow1996}T. J. Marrow, M. Aindow, P. Prangnell, M. Strangwood, J. F. Knott, Hydrogen-assisted stable crack growth in Iron-3 wt$\%$ silicon steel, Acta. mater., 44 (1996) 3125-3140
\bibitem{Chen1990}X. Chen, T. Foecke, M. Lii, Y. Katz, W. W. Gerberich, The role of stress state on hydrogen cracking in Fe-Si single crystals, Eng. Fract. Mech., 35 (1990) 997-1017
%\bibitem{Song2014}J. Song, W.A. Curtin, Acta Mater., 68 (2014) 61-69
%\bibitem{Miyamoto2009}M. Miyamoto, D. Nishijima, Y. Ueda, R.P. Doerner, H. Kurishita, M.J. Baldwin, S. Morito, K. Ono, J. Hanna, Nucl. Fusion 49 (2009) 650356.
%\bibitem{Jia2015}Y.Z. Jia, G. De Temmerman, G.N. Luo, H.Y. Xu, C. Li, B.Q. Fu, W. Liu, J. Nucl. Mater. 457 (2015) 213-219
%\bibitem{Jia2016}Y.Z. Jia, W. Liu, B. Xu, G.N. Luo, S.L. Qu, T.W. Morgan, G. De Temmerman, J. Nucl. Mater., 477 (2016) 165-171
%\bibitem{Kresse1993}G. Kresse, J. Hafner, Phys. Rev. B, 47 (1993) 558-561
%\bibitem{Kresse1996}G. Kresse, J. Furthmuller, Phys. Rev. B, 54 (1996) 11169-11186
%\bibitem{Blochl1994}P.E. Blochl, Phys. Rev. B 50 (1994) 17953.
%\bibitem{Perdew1992}J.P. Perdew, J.A. Chevary, S.H. Vosko, K.A. Jackson, M.R. Pederson, D.J. Singh, et al., Phys. Rev. B 46 (1992) 6671
%\bibitem{Perdew1993}J.P. Perdew, J.A. Chevary, S.H. Vosko, K.A. Jackson, M.R. Pederson, D.J. Singh, et al., Phys. Rev. B 48 (1993) 4978(E)
%\bibitem{Switendick1979}A. C. Switendick, Zeitschrift f¨¹r Physikalische Chemie 117 (1979) 89
%\bibitem{Becquart2009}C.S. Becquart, C. Domain, J. Nucl. Mater., 386-388 (2009) 109-111
%\bibitem{Liu2009}Y.L. Liu, Y. Zhang, G.N. Luo, G.H. Lu, J. Nucl. Mater., 390-391 (2009) 1032-1034
\end{thebibliography}

 \end{document}